\documentclass[preprint,12pt]{elsarticle}
\usepackage[utf8]{inputenc}
\usepackage{subscript}
\usepackage[compatibility=false]{caption}

\usepackage{multirow}
\usepackage{ragged2e}

\usepackage{booktabs}
\usepackage{siunitx}
\usepackage[section]{placeins}
\usepackage{xcolor,colortbl}
\usepackage[colorlinks=true, linkcolor=blue, citecolor=blue, urlcolor=blue]{hyperref}

\makeatletter
\def\ps@pprintTitle{%
  \let\@oddhead\@empty
  \let\@evenhead\@empty
  \def\@oddfoot{\reset@font\hfil\thepage\hfil}%
  \let\@evenfoot\@oddfoot
}
\makeatother
\usepackage{wrapfig}
\usepackage[a4paper, left=3cm, right=3cm, top=3cm, bottom=3cm]{geometry}
\usepackage{makecell}
\usepackage{amssymb}
\usepackage{amssymb}
\usepackage{textcomp}
\usepackage{amsmath}
\usepackage{rotating}
\usepackage{adjustbox}
\usepackage{booktabs}
\usepackage{tabularx}
\usepackage{multirow}
\usepackage{graphicx}
\usepackage{soul}
\usepackage{xcolor}
\sethlcolor{yellow}
\usepackage{caption}
\usepackage{subcaption}
\usepackage{siunitx}
\usepackage{float}
\usepackage{soul}

\usepackage[colorlinks=true,linkcolor=blue,citecolor=blue,urlcolor=blue]{hyperref}

\begin{document}

\begin{frontmatter}

\title{Electronic Structure of UGe$_{2\pm x}$ Thin Films from Photoelectron Spectroscopy}

\author[inst1,inst2]{Sonu George Alex}
\author[inst1]{Oleksandr Romanyuk}
\author[inst1]{Ivan Zorilo}
\author[inst1]{Alexander V. Andreev}
\author[inst3]{Frank Huber}
\author[inst3]{Thomas Gouder}
\author[inst4]{Petr Malinský}
\author[inst1]{Maliha Siddiqui}
\author[inst1]{Alexander B. Shick}
\author[inst1]{Evgenia A. Tereshina-Chitrova}

\address[inst1]{Institute of Physics CAS, Prague, 18200 Czech Republic}
\address[inst2]{Faculty of Mathematics and Physics, Charles University, Prague, 12116 Czech Republic}
\address[inst3]{European Commission, Joint Research Centre (JRC), Karlsruhe, DE-76125 Germany}
\address[inst4]{Nuclear Physics Institute CAS, Department of Neutron Physics, Řež, 25068 Czech Republic }

\begin{abstract}
Uranium digermanide UGe$_2$, the first ferromagnetic superconductor, represents a key composition in the U–Ge system dominated by U-5$f$ states. To examine the impact of controlled stoichiometric deviations on the electronic structure, UGe$_{2\pm x}$ thin films ($0 \le x \le 1$) were prepared by triode sputtering in an argon atmosphere and studied on pristine surfaces by X-ray (XPS) and Ultraviolet (UPS) photoelectron spectroscopy. XPS and UPS reveal a robust metallic valence band with a dominant U-5$f$ contribution at the Fermi level and a broad incoherent feature at higher binding energies, without qualitative changes in spectral line shape across the composition range. The experimental spectrum of UGe$_2$ thin films is well reproduced by DFT+U(ED) valence-band calculations combining density functional theory with exact diagonalization of the multiconfigurational U-5$f$ shell. These results demonstrate that the overall U–Ge electronic framework of UGe$_2$ thin films remains resilient to moderate stoichiometric deviations, providing a reliable electronic baseline for future studies of interface- and heterostructure-driven phenomena in uranium-based systems.


\end{abstract}


\end{frontmatter}


\section{Introduction}

Uranium intermetallics provide a compelling arena for examining electronic behaviour that arises from 5$f$ states governed by strong relativistic effects and electron–electron correlations~\cite{sech}. UGe$_2$, the first known ferromagnetic superconductor (under moderate pressure)~\cite{saxena2000,pfleiderer2002}, represents one of the most fascinating cases, owing to the interplay between its 5$f$-derived electronic bands, magnetism, and superconductivity. Accessing this interplay requires experimental probes that are directly sensitive to the occupied electronic states and their degree of itinerancy or localization. In this respect, photoemission spectroscopy (PES) plays a central role, as it provides direct information on the energy distribution and orbital character of the occupied electronic states. Previous PES experiments on UGe$_2$ show that its electronic structure is mainly governed by U-5$f$ states mixed with Ge-4$p$ and U-6$d$ orbitals, giving a metallic spectrum with a sharp peak at the Fermi level and broader spectral weight at higher binding energies of approximately $1$–$3$ eV, which cannot be reproduced within a coherent band-like description provided by standard density functional theory~\cite{samsel2011}. This limitation motivates the use of more advanced theoretical approaches when analyzing the U–Ge electronic structure.

The electronic structure of stoichiometric UGe$_2$ under pressure has been widely studied~\cite{Terashima2001, Terashima2002}, several observations indicate that its 5$f$ bands are highly sensitive to modest changes in hybridization. Pressure tuning modifies band dispersions and the ordered moment with only sub-percent structural changes~\cite{Haga2002,Oomi2002}. Valence-band PES reveals that the relative weight of coherent and incoherent 5$f$ features depends strongly on the local screening environment. Angle-resolved PES (ARPES) further shows deviations from band theory that arise from subtle details of U–ligand hybridization~\cite{fujimori2015}. Together, these results suggest that even moderate chemical perturbations may influence the U–Ge electronic structure. Yet the effect of controlled deviations from stoichiometry on the 5$f$ states of UGe$_2$ remains essentially unexplored.

Nonequilibrium thin-film growth enables such chemical tuning~\cite{tereshina2023}. By adjusting sputtering parameters, UGe$_{2\pm x}$ thin films with well-defined off-stoichiometric compositions can be reproducibly synthesized. Earlier attempts to prepare UGe$_2$ thin films by sputtering resulted in amorphous and heavily oxidized material with strongly suppressed magnetic properties~\cite{Homma1998}, underscoring the challenge of stabilizing this compound in thin-film form. The films produced in this work provide direct access to compositions inaccessible in bulk and allow systematic investigation of how stoichiometry influences the electronic structure of UGe$_{2\pm x}$. Although the majority of PES studies has naturally focused on bulk UGe$_2$~\cite{beaux2011,Fujimori2016}, Soda et al.~\cite{soda1991} and Ejima et al. \cite{Ejima1996} also examined a limited number of additional bulk Ge:U compositions. Subsequent studies clarified that compositions in this range correspond to multiphase mixtures rather than well-defined single-phase compounds \cite{Troc2002}. As a result, systematic photoemission studies of stoichiometry-controlled UGe$_{2\pm x}$ compositions are lacking in both bulk materials and thin-film form.

In this work, we study the evolution of the electronic structure in U–Ge thin films with controlled deviations from the UGe$_2$ stoichiometry and investigate the rigidity of the system against such chemical perturbations. Using \textit{in situ} X-ray photoelectron spectroscopy (XPS) and ultraviolet photoelectron spectroscopy (UPS), we analyze valence-band features and core-level spectra across the UGe${_{2\pm x}}$ ($0 \le x \le 1$) series on impurity-free surfaces. These measurements enable access to compositions that are difficult or inaccessible in bulk form and allow us to examine the sensitivity of the UGe$_2$ electronic structure to controlled compositional variations. Importantly, the thin films are used as an electronic probe and are not intended to represent equilibrium bulk phases.

To interpret the experimental spectra, we compare them with density functional theory calculations supplemented by a Hubbard $U$ and exact diagonalization (DFT+U(ED)) for bulk stoichiometric UGe$_2$, in which the multiconfigurational U-5$f$ shell is treated explicitly. For the stoichiometric composition, we additionally compute the momentum-resolved spectral function to benchmark the calculated 5$f$ band dispersions against available ARPES data \cite{fujimori2015}. We show that the U-5$f$ electronic structure of UGe$_2$ remains remarkably robust against controlled stoichiometric deviations in nonequilibrium thin films.

\renewcommand{\arraystretch}{1.6}  

\section{Materials and Methods}

Thin films of UGe$_{2\pm x}$ ($0 \le x \le 1$) were prepared using a the ultra-high vacuum (UHV) triode sputtering deposition system previously employed in preparation of various uranium-based alloys~\cite{u-te,tereshina2021,paukov2018}. The sputtering target was a single-crystal UGe$_2$ rod grown by the Czochralski method from natural uranium (99.9 wt.\% purity). Its phase purity was confirmed by powder X-ray diffraction (Bruker D8) and Laue backscattering (not shown here). The target was positioned 20 mm opposite the substrate, directly exposed to the plasma. Argon (5N purity), additionally purified with an Oxisorb® cartridge, was used as the working gas. 

In order to systematically alter the films stoichiometry, the Ar pressure and discharge current were varied while keeping the target voltage fixed at $-800~\mathrm{V}$, yielding a range of compositions (Table~1). Lower working pressure reduces scattering in the plasma and increases the arrival probability of Ge (lower mass than U) at the substrate, resulting in higher Ge:U ratios. The target current likewise controls the relative sputter yield: at low currents the deposition approaches the bulk UGe$_2$ stoichiometry, whereas increasing the current enhances preferential sputtering and re-sputtering of Ge, yielding progressively U-rich films. Under the present conditions, a working pressure of $1.08 \times 10^{-3}$~mbar and a target current of 0.8--1~mA provided films with Ge:U = 2, representing the optimal window for reproducing the near-stoichiometric composition. 

We used 10×10 mm$^2$ Si(001) wafers as the substrates, cleaned by Ar$^+$ ion bombardment. All samples for the study were prepared at room temperature to minimize interdiffusion with the substrate. Surface contaminants, such as oxygen, were effectively reduced during the early stages of deposition. Deposition rates ranged from 0.01 to 0.1 nm/s, depending on the argon pressure and target current (see supplementary section S1 for greater details) 




\begin{table}[H]
\centering
\caption{The list of UGe$_{2\pm x}$ thin films and deposition parameters. Throughout this work, compositions are expressed as atomic Ge:U ratios as directly determined from XPS analysis.}
\label{tab:deposition_parameters}

{\fontsize{10}{6}\selectfont
\renewcommand{\arraystretch}{3}

\begin{tabular}{@{} l c c c c @{}}
\toprule
\textit{\textbf{x}} &
\textbf{Ge/U} &
\textbf{Substrate} &
\textbf{Target current (mA)} &
\textbf{Argon gas pressure (mbar)} \\
\midrule
1.0   & 1.0 & Si & 2.0 & $10 \times 10^{-3}$ \\
0.7   & 1.3 & Si & 2.0 & $5 \times 10^{-3}$  \\
0.2   & 1.8 & Si & 1.2 & $1 \times 10^{-3}$  \\
0     & 2.0 & Si & 0.8 & $1 \times 10^{-3}$  \\
-0.2  & 2.2 & Si & 1.0 & $1 \times 10^{-3}$  \\
\bottomrule
\end{tabular}

\vspace{1.5mm}
}

\end{table}



After deposition, the films were immediately transferred under ultra-high-vacuum to the neighbouring analysis chamber. XPS measurements were carried out using monochromated Al K$\alpha$ radiation ($h\nu = 1486.6$ eV), and UPS study was performed with He II excitation ($h\nu = 40.81$ eV), using a SPECS PHOIBOS~150 MCD-9 hemispherical analyser. The overall energy resolution was approximately 0.5 eV for core-level XPS and about 50 meV for UPS. The energy scale was calibrated using the Au-4$f_{7/2}$ (83.9 eV) and Cu-2$p_{3/2}$ (932.7 eV) core-level positions for XPS and the Fermi edge of Au for UPS.

Survey spectra were collected to confirm the absence of extraneous elements (as shown in Supplementary Fig.~\ref{fig:xps_survey}). High-resolution U 4$f$ (400–300 eV) and Ge 3$d$ (33–27 eV) spectra were acquired for quantitative evaluation, and the corresponding stoichiometry values were used to refine the sputtering conditions (working-gas pressure and target current) for subsequent depositions. The stoichiometry was determined by taking the ratio of the integrated intensities of the U 4\textit{f} and Ge 3\textit{d} core-level lines, calculated after the Shirley background subtraction. Given the high oxygen affinity of uranium and the surface sensitivity of XPS, the presence of oxygen-related contributions was explicitly examined. The Ge 3$d$ region partially overlaps with the Ge LMM Auger structure, and the O 1$s$ core level lies close to this energy window, making direct quantification of the oxygen concentration challenging. Nevertheless, deconvolution of the Ge 3$d$ core-level spectrum enabled reliable extraction of the Ge:O ratio, as shown in Supplementary Fig.~\ref{fig:xps_ge_deconvolution}. The oxygen content was estimated to be below 4\% within the XPS information depth, while only trace levels of carbon contamination were detected in thin films. Because XPS is surface sensitive, complementary RBS measurements were performed to verify the Ge:U ratio through the full film thickness. The RBS analysis confirms the stoichiometry within ±5\% relative uncertainty, in agreement with XPS. Full RBS results will be presented in a subsequent publication on the physical properties of these films. 

\textcolor{red}{Structural characterization was performed by grazing-incidence X-ray diffraction (GIXRD). Representative diffraction patterns are shown in Fig.~S3 in the Supplementary Information. Measurements were carried out using a Rigaku SmartLab diffractometer equipped with a 9 kW Cu rotating anode source (Cu K$\alpha$, $\lambda$ = 0.15418 nm) at a constant incidence angle $\alpha_i = 0.6^\circ$.}

To place the experimental data into theoretical context, we employ the DFT+U(ED) extension ~\cite{shick2021} of the DFT+U method, which combines conventional DFT with exact diagonalization (ED) of the multiconfigurational U 5$f$ shell in UGe$_2$. This approach explicitly treats the atomic multiplet structure and thereby captures correlation effects beyond mean-field Hubbard-$U$ approximations. Very recently, DFT+U(ED) was successfully applied to the magnetic properties of UTe$_2$ in an external magnetic field~\cite{shick2024}. The calculations reproduced both the spin and orbital magnetic moments on the U atom in good agreement with x-ray magnetic circular dichroism measurements at the U $M_{4,5}$ edges~\cite{wilhelm2023}, as well as with ARPES and PES data ~\cite{shick2024}.

\section{Results and Discussion}

\subsection{Core-Level XPS Study}
Core-level photoemission data across bulk U–Ge binaries are sparse and not systematic. Besides the widely cited UGe$_2$ spectra \cite{fujimori2015,samsel2011}, the main comparative reference remains the study by Ejima et al.~\cite{Ejima1996}, who found rather similar U-4$f$ spectra for UGe$_2$ and samples then assigned to the U$_3$Ge$_4$ phase. However, the limited sample quality and the subsequent clarification that U$_3$Ge$_4$ is not a stable single-phase compound \cite{Troc2002} prevent firm conclusions regarding the evolution of U~4$f$ core-level spectra with composition. This motivates systematic studies over a broader composition range in the vicinity of UGe$_2$ on pristine surfaces.

Figure~1(a) shows the U~4$f$ core-level spectra of the UGe$_{2\pm x}$ thin films with Ge:U ratios listed in Table~1. All spectra exhibit the characteristic spin–orbit–split doublet, with a 4$f_{7/2}$–4$f_{5/2}$ separation of 10.9~eV, consistent with previous studies of uranium compounds~\cite{Ilton2011}. The U~4$f_{7/2}$ and U~4$f_{5/2}$ peak positions are typically observed at $377.3 \pm 0.1$~eV and $388.2 \pm 0.1$~eV, respectively, in good agreement with synchrotron-based PES data reported \cite{fujimori2015} for bulk UGe$_2$ (Fig. 1 (a,b)).

\begin{figure}[H]
    \centering
    \makebox[\textwidth][c]{%
    \includegraphics[width=1.3\textwidth]{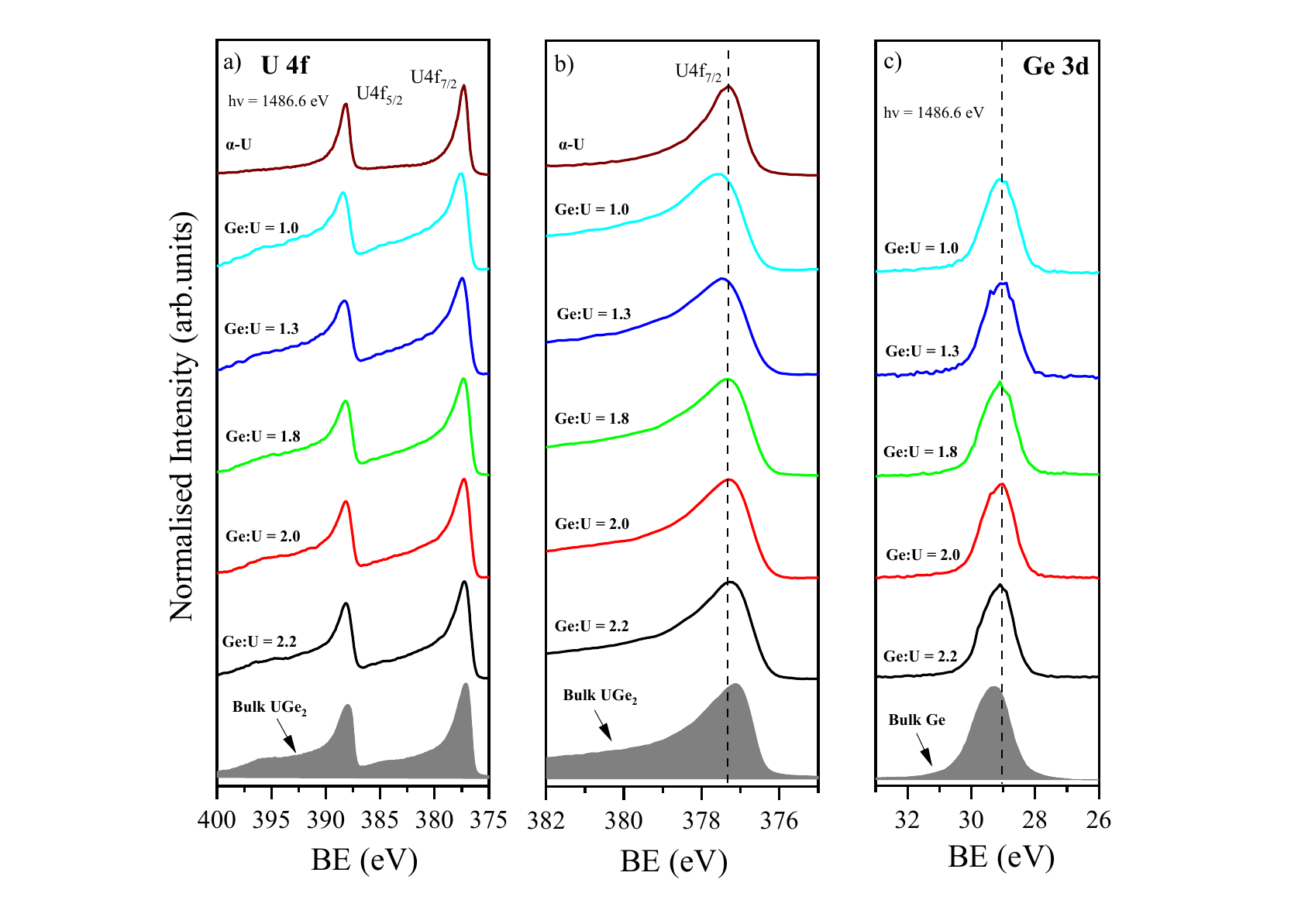}%
    }
    \caption{XPS core-level spectra of the U--Ge system measured at $h\nu = 1486.6$~eV. 
(a) U-4$f$ spectra, (b) expanded U-4$f$$_{7/2}$ region, and (c) Ge-3$d$ spectra for different Ge:U ratios, including bulk Ge. Spectra are normalized to the respective core-level maximum and vertically offset for clarity; dashed lines are guide to the eye. The bulk UGe$_2$ spectrum is a synchrotron soft–x-ray PES data adopted from \cite{fujimori2015}, measured at $h\nu$ = 800 eV and $T$ = 120 K.}
    \label{fig:xps-uge2}
    \end{figure} 
\begin{table}[htbp]
\caption{Crystallographic data and selected physical properties of bulk U--Ge compounds. The table lists the crystal structures, space group, Ge:U atomic ratio, Sommerfeld coefficient $\gamma$, magnetic ground state (PM: paramagnetic; FM: ferromagnetic), Curie temperature $T_C$ where applicable, effective magnetic moment $\mu_{\mathrm{eff}}$, and nearest-neighbour U--U and U--Ge distances.}
\centering

\renewcommand{\arraystretch}{1.6}
\setlength{\tabcolsep}{2.5pt}

{\footnotesize
\resizebox{\linewidth}{!}{%
\begin{tabular}{l
                >{\RaggedRight\arraybackslash}p{2.1cm}
                c
                c
                c
                >{\RaggedRight\arraybackslash}p{2.6cm}
                c
                c
                c}
\toprule
\textbf{Compound} &
\textbf{\shortstack{Structure\\type}} &
\textbf{\shortstack{Space\\group}} &
\textbf{\shortstack{Ge/U\\ratio}} &
\textbf{\shortstack{$\gamma$\\(mJ\,mol$_\mathrm{U}^{-1}$\,K$^{-2}$
)}} &
\textbf{Magnetism} &
\textbf{\shortstack{$T_C$\\(K)}} &
\textbf{\shortstack{$\mu_{\mathrm{eff}}$\\($\mu_B$)}} &
\textbf{\shortstack{$d_{\mathrm{U-U}}/d_{\mathrm{U-Ge}}$\\(\AA)}} \\
\midrule

U$_5$Ge$_4$ &
Ti$_5$Ga$_4$ (hex.) &
$P6_3/mcm$ &
0.8 &
53 \cite{Pikul2014} &
Pauli PM \cite{Boulet1997U5Ge4} &
-- &
-- &
2.93 / 2.87 \\

UGe &
ThIn (orth.) &
$Pbcm$ &
1.0 &
137 \cite{Pikul2014} &
Curie--Weiss PM \cite{Troc2002} &
-- &
2.36 \cite{Troc2002} &
2.92 / 2.82 \\

U$_3$Ge$_5$ &
AlB$_2$ (hex.) &
$P6/mmm$ &
$\sim$1.67 &
48 \cite{Pikul2014} &
FM \cite{Boulet1999U3Ge5} &
94 &
2.56 \cite{Troc2002} &
3.95 / 2.44 \\

UGe$_2$ &
ZrGa$_2$ (orth.) &
$Cmmm$ &
2.0 &
30--33 \cite{Onuki1992,Troc2012_UGe2_Dualism} &
FM \cite{boulet1997} &
52 &
2.9 \cite{deVisser2010EMSAT} &
3.84 / 2.91 \\

UGe$_3$ &
AuCu$_3$ (cub.) &
$Pm\overline{3}m$ &
3.0 &
20.4 \cite{vanMaaren1974} &
Pauli PM \cite{Onuki2014Fermi} &
-- &
-- &
4.19 / 2.97 \\

\bottomrule
\end{tabular}%

}}
\vspace{2mm}

{\footnotesize
\justifying
\textbf{Note:} Early studies reported additional compounds U$_7$Ge, U$_5$Ge$_3$~\cite{Onuki1990PhysicaB} and U$_3$Ge$_4$ ~\cite{soda1991}. Later works \cite{Troc2002,Boulet1997U5Ge4} showed that these assignments do not correspond to stable equilibrium phases; instead, they represent multiphase mixtures within the U–Ge phase diagram.

}
\end{table}

The U-4$f$ spectra of the UGe$_{2\pm x}$ thin films shown in Fig.~1 (a,b) exhibit strongly asymmetric main lines accompanied by higher-binding-energy shake-up satellites, including a structure at 383–384~eV adjacent to the U-4$f_{7/2}$ peak and a weak hump at 395–396~eV on the high-binding-energy side of the U-4$f_{5/2}$ component. Typically, the overall spectral shape in uranium and actinide compounds reflects a balance between atomic multiplet effects and the efficiency of core-hole screening in the photoemission final state~\cite{fujimori2015,gouder2002,Thompson2024}. Well-screened final states give rise to relatively sharp main peaks, as observed in $\alpha$-U (Fig. 1(a,b)). In contrast, less efficiently screened final states lead to additional spectral weight at higher binding energy and an apparent broadening of the peaks. Consistent with this understanding, more itinerant U–5$f$ systems (e.g. UB$_2$) tend to exhibit mostly sharp main lines with weaker satellites, whereas strongly correlated compounds (such as UGe$_2$, UCoGe, etc.) display more intense satellite features alongside an asymmetric main line \cite{fujimori2015}.

\begin{figure}
        \centering
    \includegraphics[width= 0.8 \textwidth]{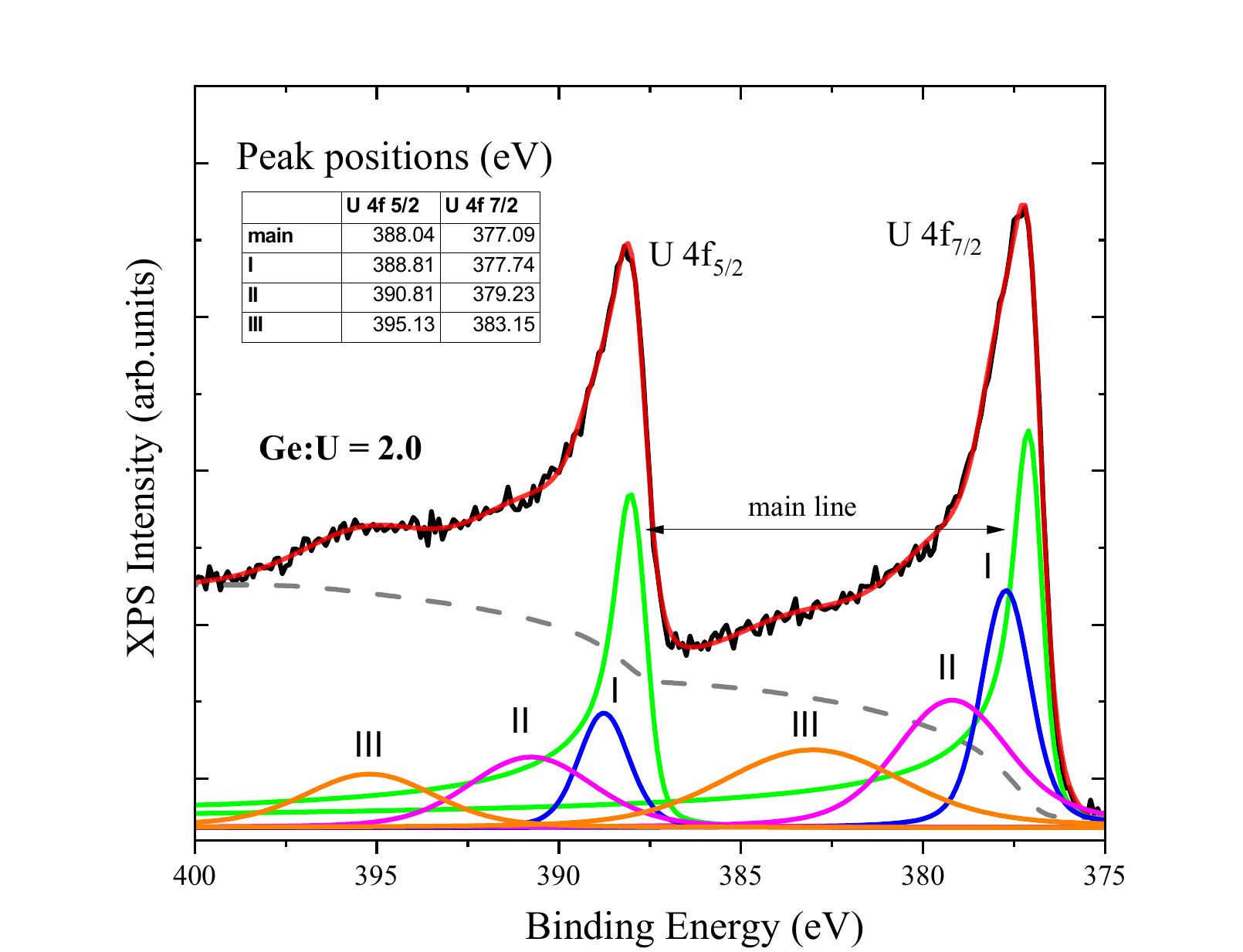} 
\caption{XPS U 4$f$ core-level spectrum of the UGe$_2$ thin film shown together with a representative phenomenological fit, in which the U-4$f$$_{7/2}$ and U-4$f$$_{5/2}$ main components are described by asymmetric Doniach–Šunjić line shapes and the satellite features (I–III) by Lorentzian–Gaussian functions. The binding energies of the main lines and satellites are indicated in eV, and the Shirley background is shown as a broken line.}
    \label{fig:core-levels_qq}
\end{figure}

To illustrate this behavior, a representative phenomenological fit of the U-4$f$ core-level spectrum is shown in Fig.~2 for the near-stoichiometric UGe$_2$ thin film (the detailed study of the spectra of other compositions of the UGe$_{2\pm x}$ series are shown in Supplement S3). The main U-4$f_{7/2}$ and U-4$f_{5/2}$ peaks were modeled using asymmetric Doniach–Šunjić functions, while Lorentzian–Gaussian functions were employed to account for the higher-binding-energy intensity, following fitting approach previously applied to bulk UGe$_2$~\cite{samsel2011}. The resulting spectral components show very good qualitative agreement with those reported for bulk UGe$_2$~\cite{samsel2011}, indicating that the essential core-level electronic structure is well preserved in the thin films and supporting their use as model systems for investigating the electronic structure of the U–Ge system.

Despite substantial differences in crystal structure across the U–Ge system, the overall shape of the U~4$f$ core-level spectra of the UGe$_{2\pm x}$ thin films remains very similar. Relative to $\alpha$-U, the U~4$f_{7/2}$ peak position is shifted to higher binding energy (Fig.~1(b)), while the Ge~3$d$ core level, which is also spin–orbit split (see details in Supplementary part, Figs. S3 and S4), is shifted to lower binding energy with respect to elemental Ge (Fig.~1(c)). The opposite sign of these shifts is consistent with charge redistribution upon compound formation, given the higher electronegativity of Ge compared to U. Between the different U–Ge compounds, the binding-energy variations of both U~4$f$ and Ge~3$d$ remain small, on the order of $\sim$0.1~eV, demonstrating that deviations from UGe$_2$ stoichiometry do not substantially affect the local electronic environment of UGe$_{2\pm x}$ thin films, in agreement with the valence-band photoemission results discussed below.

\subsection{Valence Band Study by XPS and UPS}
The valence-band spectra of the UGe$_{2\pm x}$ thin films were measured using a combination of XPS and UPS techniques, which provide complementary sensitivity to different regions of the occupied states. While XPS probes the full valence manifold with higher bulk sensitivity, UPS offers a higher photoionization cross-section for U 5$f$ states and superior resolution near the Fermi level E$_F$ \cite{opeil2006} (see Table~4). 

\begin{table}[ht]
\centering
\caption{Ratio of atomic photoionization cross sections per electron to that of the U
$5f$ states for XPS at 1486.6~eV (Al K$\alpha$) and UPS at 40.8~eV (He~II) (calculated from Ref.~\cite{YehLindau1985}).}

\label{tab:cross_section_ratio}

\small
\setlength{\tabcolsep}{10pt}      
\renewcommand{\arraystretch}{1.15} 

\begin{tabular}{lcc}
\toprule
\multirow{3}{*}{\textbf{Orbital}} &
\multicolumn{2}{c}{\textbf{Cross-Section Ratio}} \\
\cmidrule(lr){2-3}
 & \textbf{XPS} & \textbf{UPS} \\
 & {\footnotesize (Al K$\alpha$, 1486.6 eV)}
 & {\footnotesize (He~II, 40.8 eV)} \\
\midrule
Ge $4s$ & 0.17  & 0.05 \\
Ge $4p$ & 0.13  & 0.03 \\
U $6d$  & 0.017 & 0.33 \\
\bottomrule
\end{tabular}
\end{table}

Figure~3(a) shows valence-band spectra acquired by XPS for films of various stoichiometry, compared with data for $\alpha$-U measured under the same conditions. All UGe$_{2\pm x}$ films exhibit finite spectral intensity at E$_F$, demonstrating their metallic character. The $\alpha$-U spectrum exhibits the characteristic triangular valence-band line shape, with a gradual increase in intensity toward $E_F$. In contrast, the UGe$_{2\pm x}$ films show a noticeably steeper rise of the near-$E_F$ emission on the high-binding-energy side. The same trend is observed in the UPS data (Fig.~3(b)), which primarily probe U~5$f$-derived states close to $E_F$~\cite{opeil2006}, indicating a modification of the near-$E_F$ U-5$f$ spectral distribution compared with $\alpha$-U.

 \begin{figure}[H]
    \centering
    \makebox[\textwidth][c]{%
        \includegraphics[width=1.2\textwidth]{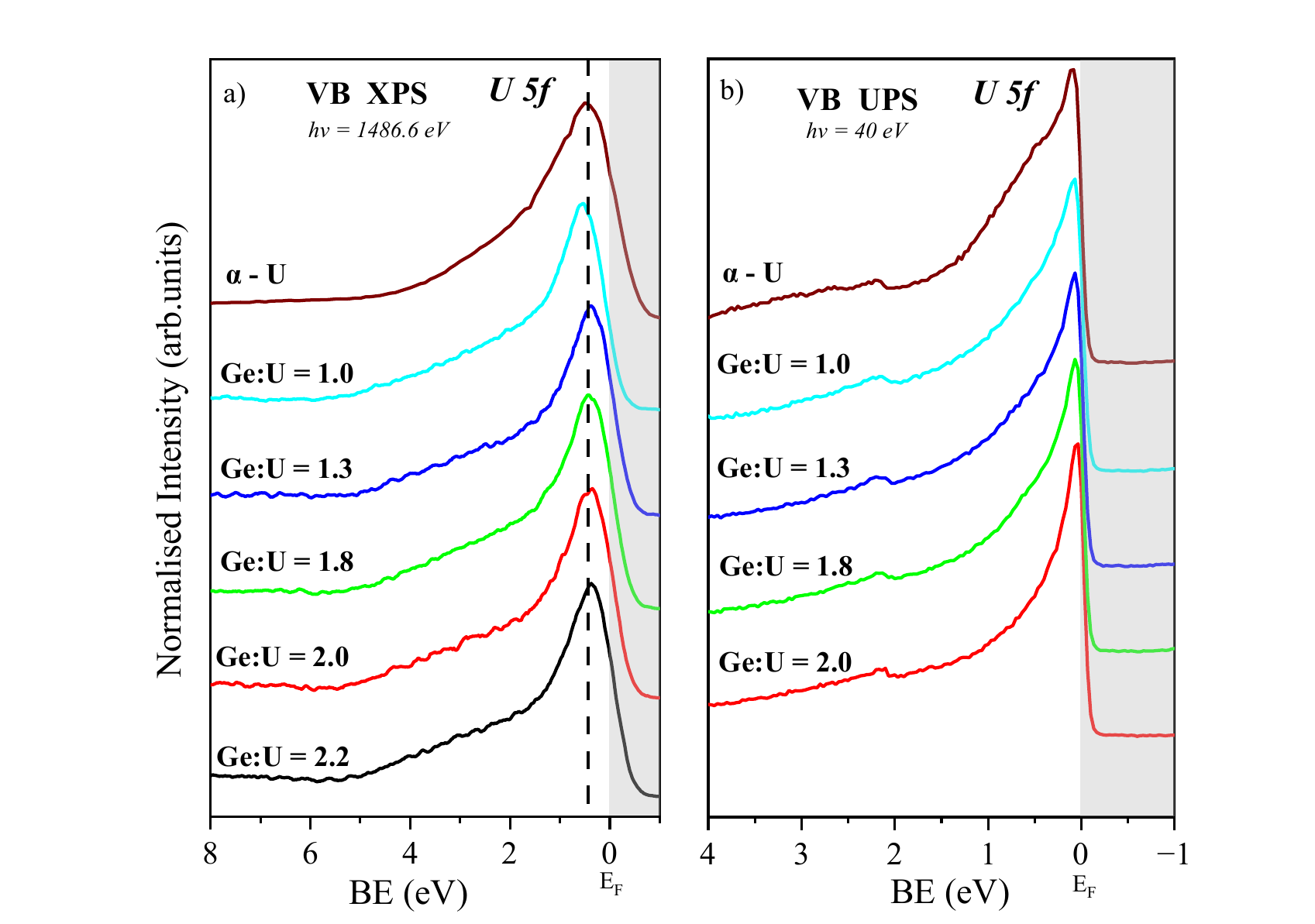}}
\caption{Valence-band spectra of $\alpha$-U and UGe$_{2\pm x}$ thin films with varying Ge:U stoichiometry measured by (a) XPS ($h\nu = 1486.6$~eV) and (b) UPS ($h\nu = 40$~eV). Spectra are normalized and vertically offset for clarity. The Fermi level ($E_F = 0$) is indicated; the shaded region denotes unoccupied states. The vertical dashed line serves as a guide to the eye.}
    \label{fig:xps-uge2_VB}
\end{figure}

At higher binding energies, the XPS spectra of the UGe$_{2\pm x}$ films display an extended tail forming a broad hump between approximately 2 and 4~eV. In $\alpha$-U, states with significant U-6$d$ character are known to occur near $\sim 2$~eV and contribute to a shoulder-like feature in UPS spectra~\cite{Reihl1985,Molodtsov1998}. Upon alloying with Ge, these U-6$d$ states strongly hybridize with Ge-4$p$ states in the same energy range \cite{samsel2011}. 

Previous LSDA/GGA band-structure calculations for bulk UGe$_2$ within a purely itinerant framework~\cite{samsel2011} reproduce both the dominant U-5$f$ contribution at the Fermi level and the hybridized U-6$d$-Ge-4$p$ bands at higher binding energies. However, the calculated density of states in this energy range remains relatively weak and does not account for the pronounced hump observed in the XPS spectra. In contrast, the calculations presented here, which explicitly include the atomic multiplet structure of the U-5$f$ shell, show that a substantial fraction of U-5$f$ spectral weight contribute significant spectral weight at higher binding energies, naturally accounting for the broad hump observed in the XPS spectra of UGe$_{2}$ and the corresponding UGe$_{2\pm x}$ films.

Across the investigated composition range, the overall valence-band line shape of the UGe$_{2\pm x}$ films remains remarkably similar. With the exception of the Ge:U = 1 film, no systematic shift of the near-$E_F$ emission is resolved within experimental sensitivity. The Ge:U = 1 film exhibits a shift of the near-$E_F$ maximum toward higher binding energy despite showing the largest Sommerfeld coefficient among the studied films (Table~2). This combination of reduced spectral weight at $E_F$ and enhanced electronic specific heat is compatible with semimetal-like behavior reported for bulk UGe~\cite{Troc2002}. Overall, these observations indicate that variations in composition primarily affect the spectral weight distribution near the Fermi level, while the characteristic valence-band line shape remains largely preserved across different U–Ge stoichiometries.

\label{tab:rbs-results}


\subsection{Theoretical calculations}

Since the discovery of superconductivity in ferromagnetic UGe$_2$ \cite{saxena2000}, it has become evident that standard density functional theory (DFT), when employed within the local density approximation (LDA) or generalized gradient approximation (GGA), is insufficient to describe the magnetic properties arising from the 5$f$ electrons. This shortcoming has motivated the development and application of extensions such as DFT+Hubbard-$U$ (DFT+U)~\cite{shick2001,shick2004,yaresko2005} and DFT with orbital polarization corrections (DFT+OPC)~\cite{eriksson1989,samsel2011,palacio2016}. However, neither approach succeeds in providing a satisfactory description of the PES data of uranium-based materials.

Motivated by these limitations of standard DFT and its extensions, we turn to the DFT+U(ED) method; the corresponding ground-state quantities for UGe$_2$ are listed in Table~4, which summarizes the 5$f$ electron count ($n_{e}^{5f}$), spin ($\mu_S$) and orbital ($\mu_L$) moments, the ratio $-\mu_L/\mu_S$, the total 5$f$-shell moment $\mu{\mathrm{tot}}=\mu_S+\mu_L$, and the total magnetic moment per formula unit $M_{\mathrm{tot}}$ (in $\mu_B$) for bulk UGe$_2$. In these calculations we fix the exchange parameter at $J=0.51$~eV and treat the Coulomb interaction $U$ as an adjustable parameter.

 \begin{table}[!h]
\centering    
\caption{5$f$ electron count ($n_e^{5f}$), spin ($\mu_S$) and orbital ($\mu_L$) moments,
the ratio  $-\mu_L/\mu_S$, and total magnetic moment $\mu_{tot}$ for the
5$f$ shell of U atom, and the total $M_{tot}$ moment per f.u.\ in $\mu_B$
units, compared with experimental data from Refs. \cite{boulet1997,kernavanois2001}.}
\begin{tabular}{ccccccc}
\hline
$U=$ & $n_e^{5f}$ & $\mu_S$ & $\mu_L$ & $-\mu_L/\mu_S$ & $\mu_{tot}$ & $M_{tot}$ \\ 
\hline
1.0 eV & 2.73 & -1.22 & 2.74 & 2.25 & 1.52 & 1.41\\
1.5 eV & 2.73 & -1.23 & 2.93 & 2.39 & 1.70 & 1.60\\
2.0 eV & 2.73 & -1.27 & 3.09 & 2.47 & 1.82 & 1.71\\
\hline
Exp.~\cite{boulet1997,kernavanois2001} &  &  &  & 2.24--2.60 &  & 1.42,1.50\\
\end{tabular}
\label{tab:table1}
\end{table}

\begin{figure}[!htbp]
\centerline{\includegraphics[angle=0,width=0.9\columnwidth,clip]{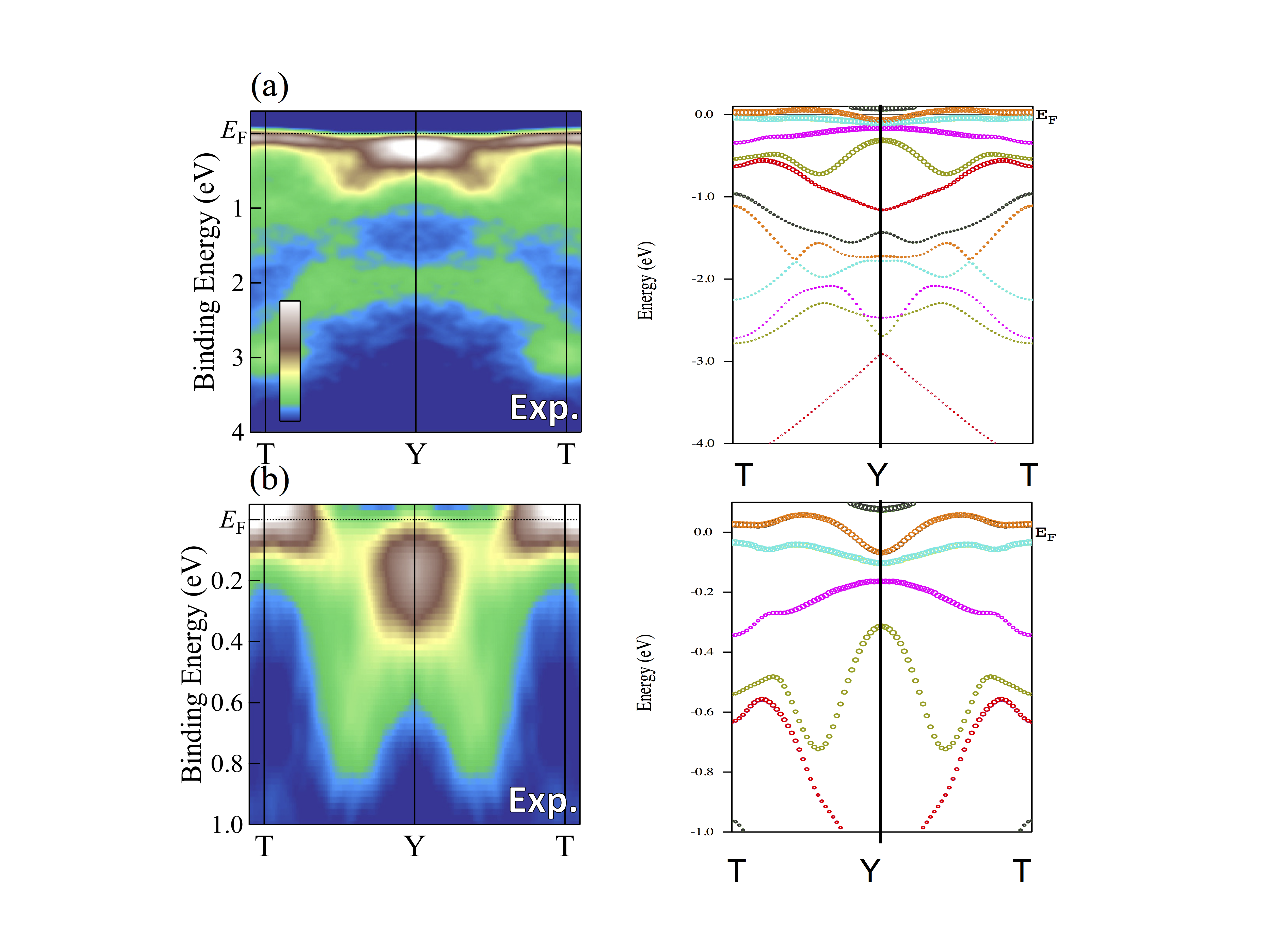}}
\caption{(a) Experimental ARPES spectra adopted from~\cite{fujimori2015} measured along the T--Y--T high-symmetry line (left), together with the calculated band structure and simulated ARPES intensity (right). 
(b) Enlarged view of the ARPES spectra (left) and corresponding calculated bands (right). 
Colors distinguish different bands, while the circle size reflects the 5$f$ character of the eigenstates.}
\label{fig:bands}
\end{figure}

 \begin{table}[!h]
\centering
\caption{$n_f$ occupation, 
non-zero probabilities $P_{f^n}$ to find the atomic eigenstates with integer occupation $f^{n}$ in the ground state of non-magnetic UGe$_2$,}
\label{tab:table2}
\begin{tabular}{cccc}
\hline
 $n_{f}$ & $P_{f^2}$ & $P_{f^3}$ & $P_{f^4}$\\
\hline
  2.7 & 0.27 & 0.49 & 0.17\\
\hline
 \end{tabular}
\end{table}
For $U$ values in the range 1–1.5~eV, the calculated ratio $-\mu_L/\mu_S$ and the total magnetic moment per formula unit show good agreement with experimental data. We further compare our results with the experimental ARPES spectra reported in Ref.~\cite{fujimori2015} (see Fig.~\ref{fig:bands}). As the measurements were carried out in the paramagnetic phase, we performed non-magnetic DFT+$U$(ED) calculations with $U = 1$~eV and $J = 0.51$~eV. 

\begin{figure}[!htbp]
\centering
\includegraphics[width=0.48\columnwidth]{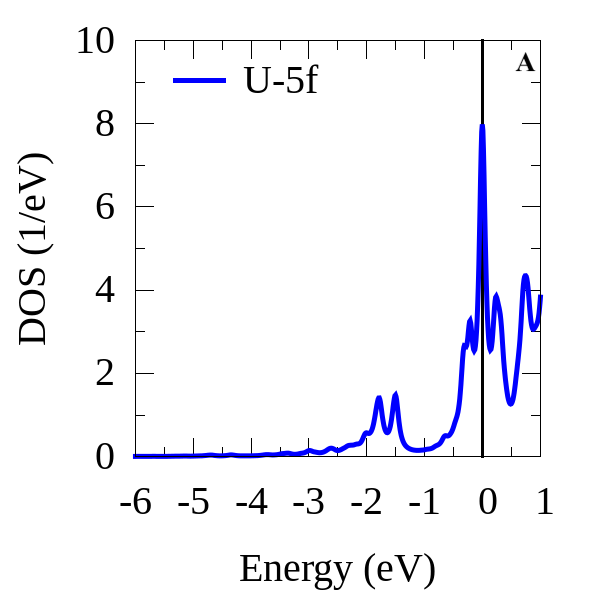}
\hfill
\includegraphics[width=0.48\columnwidth]{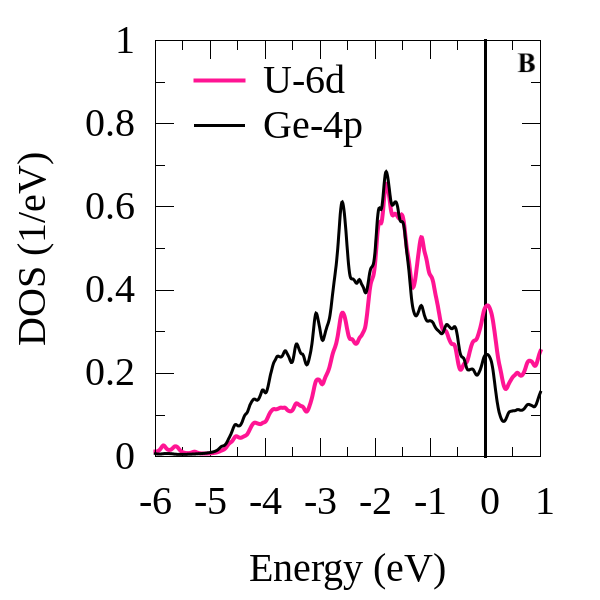}
\caption{Calculated projected density of states for UGe$_2$: 
(a) U-5$f$ states; \textcolor{red}{(b) U-6$d$ and Ge-4$p$ states.}}
\label{fig:dos}
\end{figure}

This comparison shows that the main features observed in the ARPES spectra are reproduced reasonably well by the DFT+$U$(ED) band structure. Moreover, the DFT+$U$(ED) band structure is in substantially closer agreement with the ARPES data than the LDA bands reported in Ref.~\cite{fujimori2015}. The corresponding ground state (GS) of the Anderson impurity model is a Kramers-degenerate doublet with $S$=-1.23, $L$=5.43, $J$=4.48. For the GS, the probabilities to find the atomic eigenstates with integer occupation $f^n$, $P_{f^2}=0.27$, $P_{f^3}=0.49$, $P_{f^4}=0.17$ (see Table~5). It illustrates that the GS can be taken as a superposition of mainly $f^2$ and $f^3$ many-body states. 
 
The 5$f$-projected density of states (DOS) is shown in Fig.5(a). Pronounced DOS peaks appear in the immediate vicinity of $E_F$, consistent with the band structure in ARPES measurements shown in Fig.~\ref{fig:dos}. Such features are reminiscent of those observed in intermediate-valence actinide systems, for example in Pu-based materials~\cite{shick2013}. We note that the LDA calculations reported in Ref.~\cite{fujimori2015} yield a broader peak at $E_F$, with an additional shoulder extending over the 0.5~eV energy range, which is not observed in our calculations. Instead, we find additional peaks distributed over an energy interval of approximately 2~eV, corresponding to partially localized multiplet transitions.  \textcolor{red}{In contrast, the U-6$d$ and Ge-4$p$ states (Fig.~5(b)) form broader hybridized bands extending between approximately -4 and -1 eV. While these bands contribute spectral weight in the same energy range as the higher-binding-energy tail observed in the XPS spectra (Fig. 3(a)), their calculated intensity remains relatively modest.}

\section{Discussion}

Based on the study of stoichiometry variations in the UGe$_{2\pm x}$ thin films, the behavior of U--Ge compounds can be placed in a broader context by comparison with other uranium compounds formed with other $p$-block ligands, such as Si, Ga, and Te. First of all, a clear contrast emerges between uranium tellurides and the lighter-ligand systems. As tellurium is known to adopt multiple bonding motifs and coordination environments~\cite{Steinberg2025Tellurides}, uranium tellurides exhibit pronounced spectral shifts, gap formation, and substantial redistribution of valence-band spectral weight with composition~\cite{u-te}. By contrast, ligands such as Si, Ga, and Ge form more rigid bonding frameworks, and changes in composition primarily modulate spectral weight rather than inducing qualitative changes in the valence-band line shape.

At this point, we emphasize that the present conclusions are based on spectroscopic probes of the local electronic structure in the UGe$_{2\pm x}$ thin films. While some degree of disorder or phase admixture cannot be excluded in sputter-deposited films, XPS and UPS primarily probe short-range chemical bonding and local electronic states and remain interpretable even when disorder suppresses fine spectral details associated with long-range order~\cite{Ley1973}. Early XPS studies by Ejima et al.~\cite{Ejima1996} showed that the U~4$f$ core-level spectra of bulk UGe$_2$ and samples assigned to U$_3$Ge$_4$ are nearly identical. The same result was found in the comparative valence band studies of bulk UGe$_2$, U$_3$Ge$_4$ and U$_5$Ge$_3$ \cite{soda1991}. The present thin-film data are fully consistent with these bulk observations.

The rigid compositional behavior is also well documented for uranium silicides and gallides. Photoelectron spectroscopy studies of U–Si bulk compounds \cite{sarma} and UGa$_x$ thin films \cite{ladiaUGa} show valence bands dominated by a strong U~$5f$-derived peak at the Fermi level, with only weak additional spectral weight at higher binding energies that has been attributed mainly to hybridized U~$d$ and ligand~$p$ states rather than to the extended U~$5f$ manifold. Within itinerant-$5f$ description \cite{ladiaUGa}, no pronounced extension of U~$5f$ spectral weight to higher binding energies is expected, consistent with the experimentally observed rigidity of the valence-band line shape.

In this context, the present U--Ge results provide an important extension. Experimentally, the valence-band spectra of U--Ge thin films closely resemble those of silicides and gallides, displaying a robust near-$E_F$ U~$5f$ peak and only modest changes with stoichiometry. However, our DFT+$U$(ED) calculations for stoichiometric UGe$_2$, which explicitly treat the atomic multiplet structure of the U~$5f$ shell and go beyond purely itinerant descriptions, reveal that a finite fraction of the U~$5f$ spectral weight contributes to binding energies at $\sim 2$ eV. Together with the smaller contributions from U-6$d$ and Ge-4$p$ states, this naturally accounts for the broad high-binding-energy hump observed in XPS.

The consistency between the calculated momentum-resolved spectral function (Fig. 4) and available experimental ARPES data \cite{fujimori2015} further supports this interpretation. ARPES confirms the presence of well-defined near-$E_F$ dispersive features with dominant U~$5f$ character, while deviations from simple band theory point to correlation effects that redistribute part of the $5f$ spectral weight away from the Fermi level without destroying the metallic character. Taken together, these results place U--Ge compounds in an intermediate regime: more rigid than uranium tellurides, yet exhibiting correlation-driven spectral weight redistribution beyond what is captured by purely itinerant $5f$ descriptions commonly applied to silicides and gallides.

\section{Conclusions}

Our study proved the feasibility of preparation of UGe$_{2\pm x}$ thin films by sputter deposition from a UGe$_2$ single-crystal target under varied conditions. By systematically adjusting the argon pressure and sputtering current, we obtained compositions spanning both U-rich and Ge-rich regimes. Core-level XPS analysis confirmed that the local uranium environment remains predominantly UGe$_2$-like across the investigated films.

Valence band XPS and UPS measurements reveal a pronounced U-5$f$ spectral weight at the Fermi level for all samples, consistent with metallic character of the films, in agreement with previous studies on bulk UGe$_2$ and bulk compounds of the U--Ge series. The broad valence-band features observed between 2–4 eV have long been recognized as insufficiently described by hybridized U-6$d$ and Ge-4$p$ states alone and are shown here to include a substantial contribution from the U-5$f$ states. The ability to reproducibly access stoichiometry-controlled UGe$_{2\pm x}$ thin films establishes a platform for future studies of interface- and heterostructure-driven phenomena in uranium-based quantum materials.


\section*{Acknowledgments}
We thank Dr. Ladislav Havela for useful discussions. This work was supported by the Czech Science Foundation (GACR) under grant no. 22-19416S. Sample fabrication was carried out within the framework of the EARL project of the European Commission Joint Research Centre, ITU Karlsruhe. Physical property measurements were conducted at the Materials Growth and Measurement Laboratory (http://mgml.eu/), supported by the Czech Research Infrastructures program (project no. LM2023065). S.G.A. acknowledges support from the Grant Agency of Charles University (GAUK, project no. 416925, starting in 2025), which supports his ongoing contributions to data analysis and interpretation. A.B.S. acknowledges partial support provided by the Czech Science Foundation (GACR) Grant No. 24-11992S, which covered the theoretical calculations.

\bibliography{references}



\renewcommand\thesection{S\arabic{section}}
\renewcommand\thefigure{S\arabic{figure}}
\renewcommand\thetable{S\arabic{table}}
\renewcommand\theequation{S\arabic{equation}}

\setcounter{section}{0}
\setcounter{figure}{0}
\setcounter{table}{0}
\setcounter{equation}{0}

\section*{Supplementary Information}
\addcontentsline{toc}{section}{Supplementary Information}
\vspace{0.2cm}

\section{Films synthesis route}
\label{sec:synthesis route}

\begin{figure}[H]
    \centering
    \begin{subfigure}{\textwidth}
        \centering
        \includegraphics[width=0.7\textwidth]{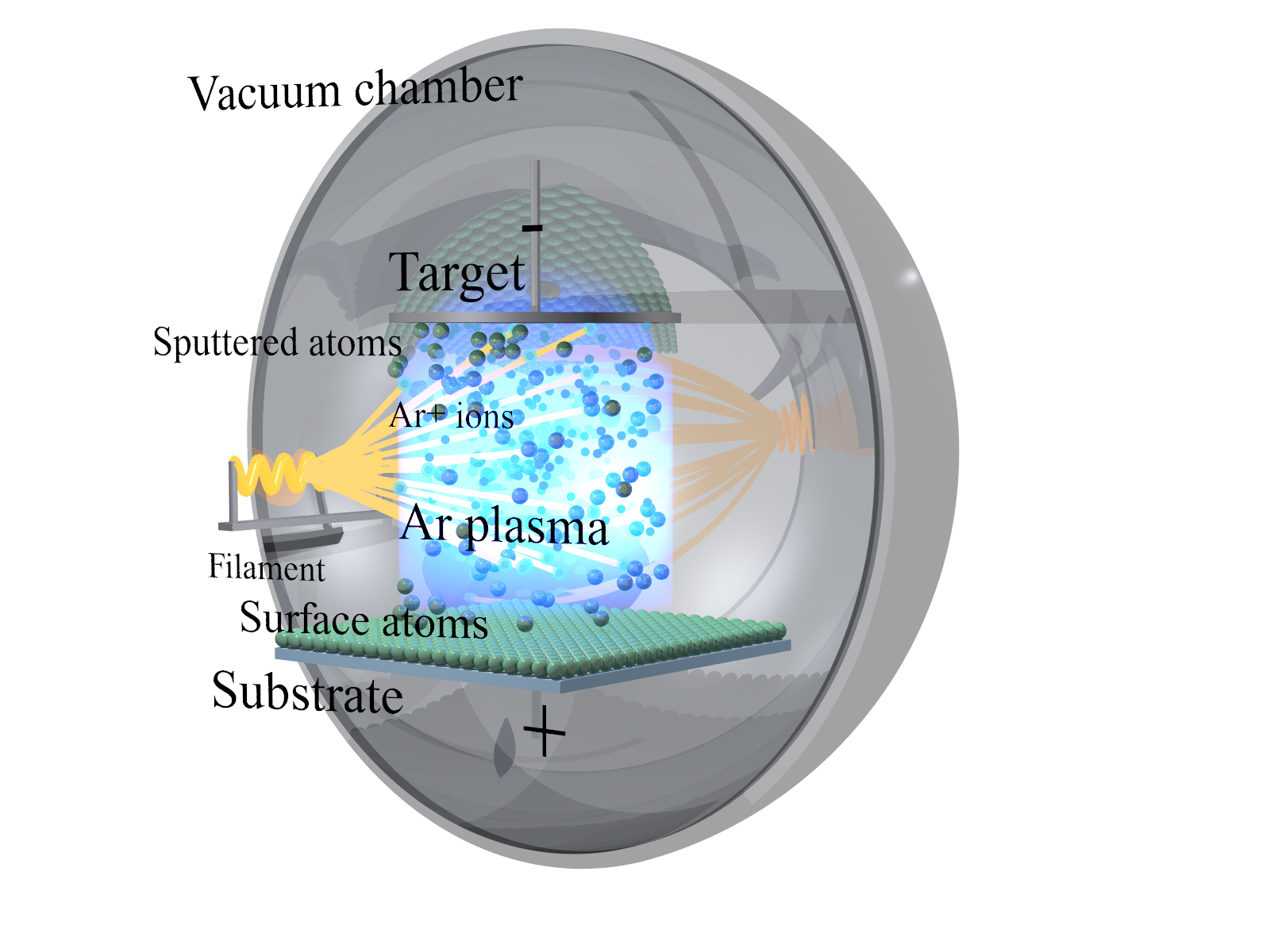}
        \caption{Schematic representation of the triode sputtering system.}
        \label{fig:triode_sputtering}
    \end{subfigure}

    \vspace{0.5cm} 

    \begin{subfigure}{\textwidth}
        \centering
        \includegraphics[width=0.8\textwidth]{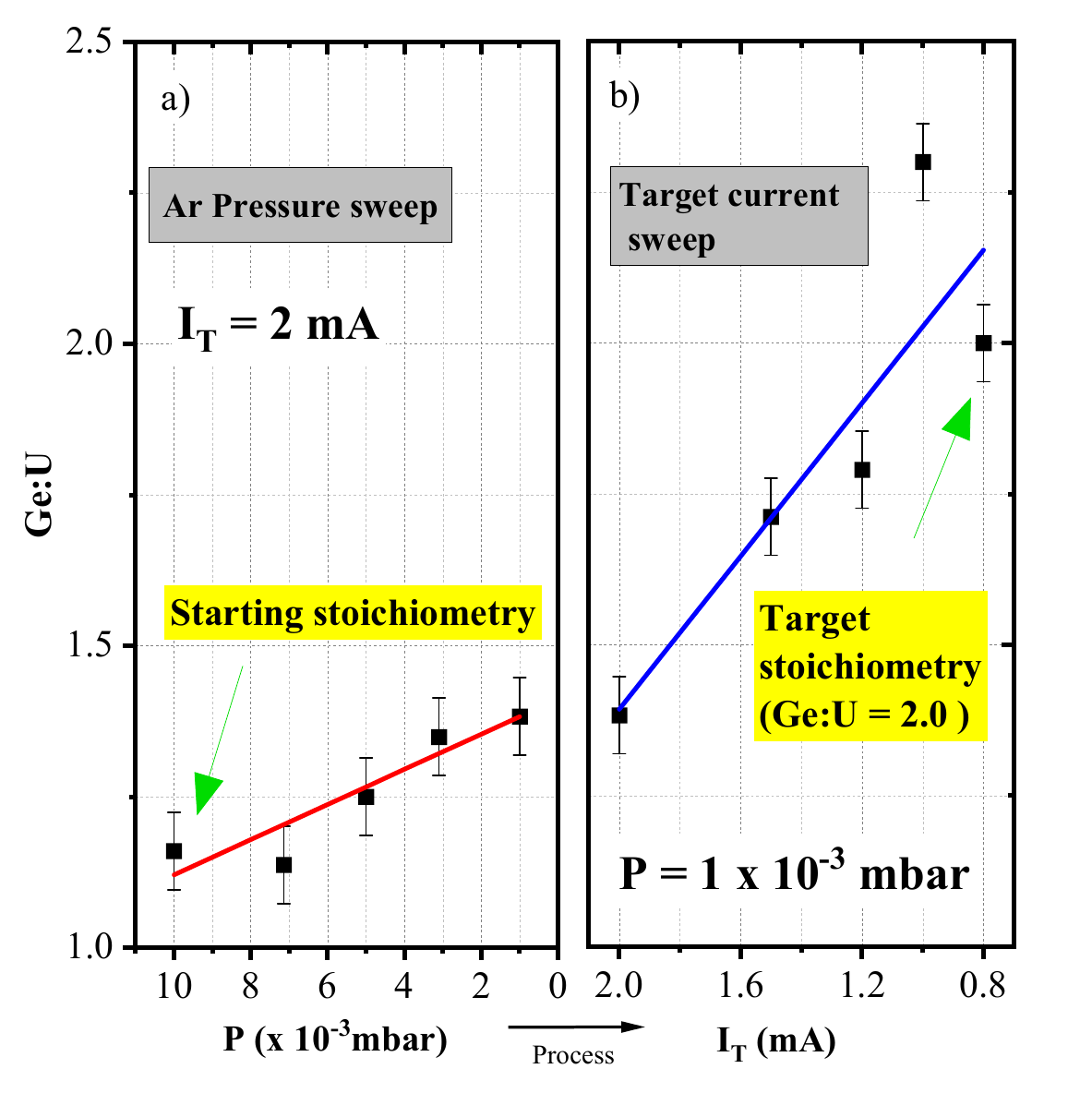}
        \caption{Dependence of the Ge:U atomic ratio on (a) argon pressure and (b) target current.}
        \label{fig:uge_ratio}
    \end{subfigure}

    \caption{Control of U--Ge films composition using triode sputtering deposition parameters.}
    \label{fig:deposition_parameters}
\end{figure}
The Ge–U thin films were deposited by triode sputtering (Fig.~\ref{fig:triode_sputtering}), where a hot filament acts as an auxiliary electron source to sustain the Ar plasma independently of the target bias. This geometry ensures a stable, dense plasma over a wide range of Ar pressures and target currents, enabling reliable deposition under low-pressure and low-current conditions.

The filament-assisted configuration decouples plasma generation from the sputtering process, allowing independent control of deposition parameters. As a result, the Ge:U ratio can be reproducibly tuned by varying either the Ar pressure at fixed target current or the target current at fixed pressure (Fig.~\ref{fig:uge_ratio}), approaching the target stoichiometry at lower currents. This highlights the key advantage of triode sputtering: precise and flexible control of thin-film composition.





\section{XPS study - survey spectra}

\begin{figure}[!htbp]
    \centering
    \includegraphics[width=0.6\textwidth]{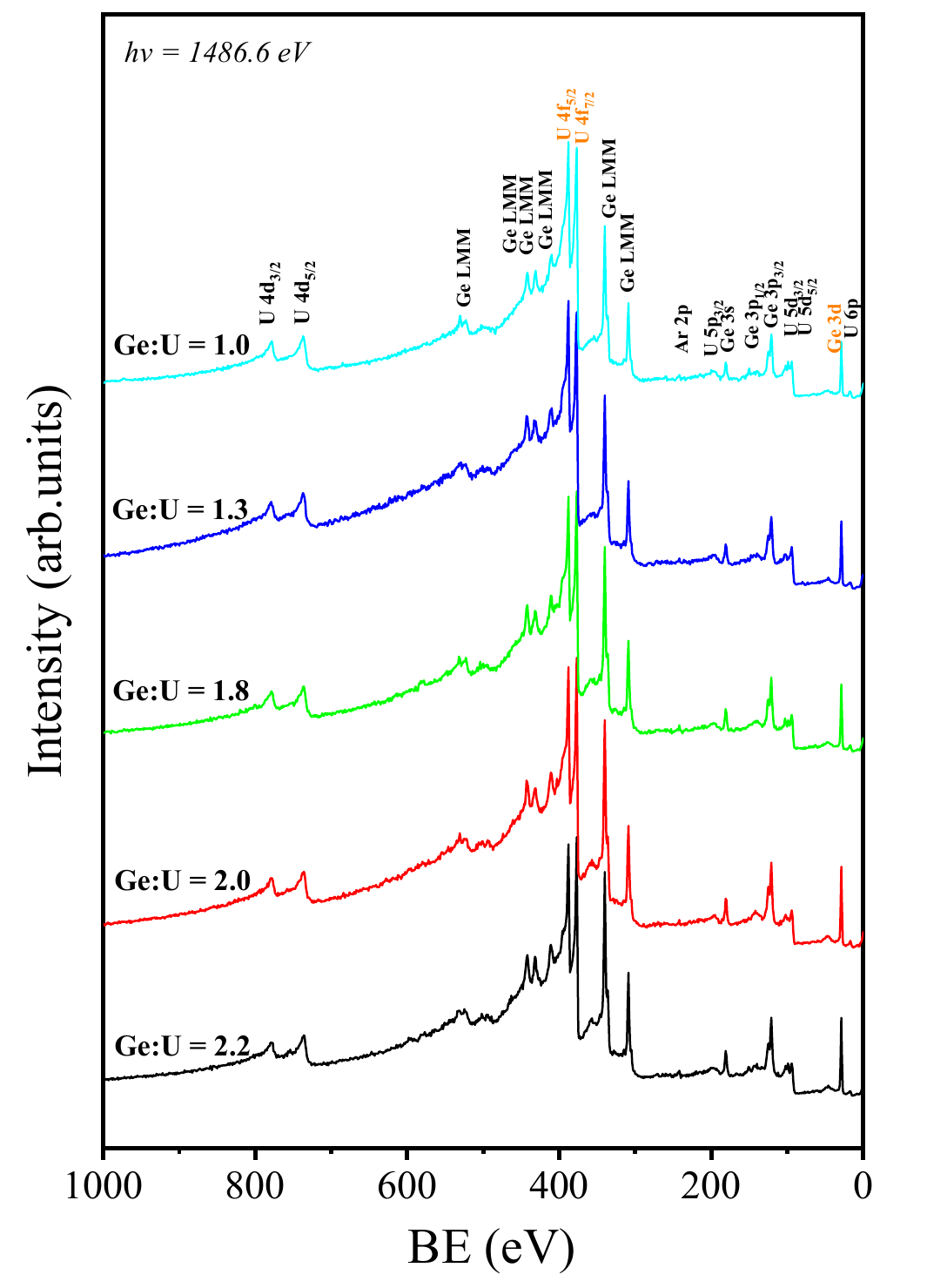}
    \caption{XPS survey spectra of UGe$_{2\pm x}$ thin films with varying Ge:U ratios recorded using Al K$\alpha$ radiation ($h\nu = 1486.6$~eV). The systematic increase in the Ge-related peak intensities with increasing Ge:U ratio demonstrates effective compositional tuning. The spectra are vertically offset for clarity.}
    \label{fig:xps_survey}
\end{figure}




\FloatBarrier
\clearpage
\section{X-ray characterization of selected films}
\FloatBarrier

Figure S3 shows representative GIXRD patterns for several UGe$_{2\pm x}$ films compositions. For other compositions within the UGe$_{2\pm x}$ series, the diffraction patterns indicated predominantly amorphous structure without clearly resolved Bragg reflections; these samples are therefore not shown.

\begin{figure}[!htbp]
    \centering
    \includegraphics[width=0.6\textwidth]{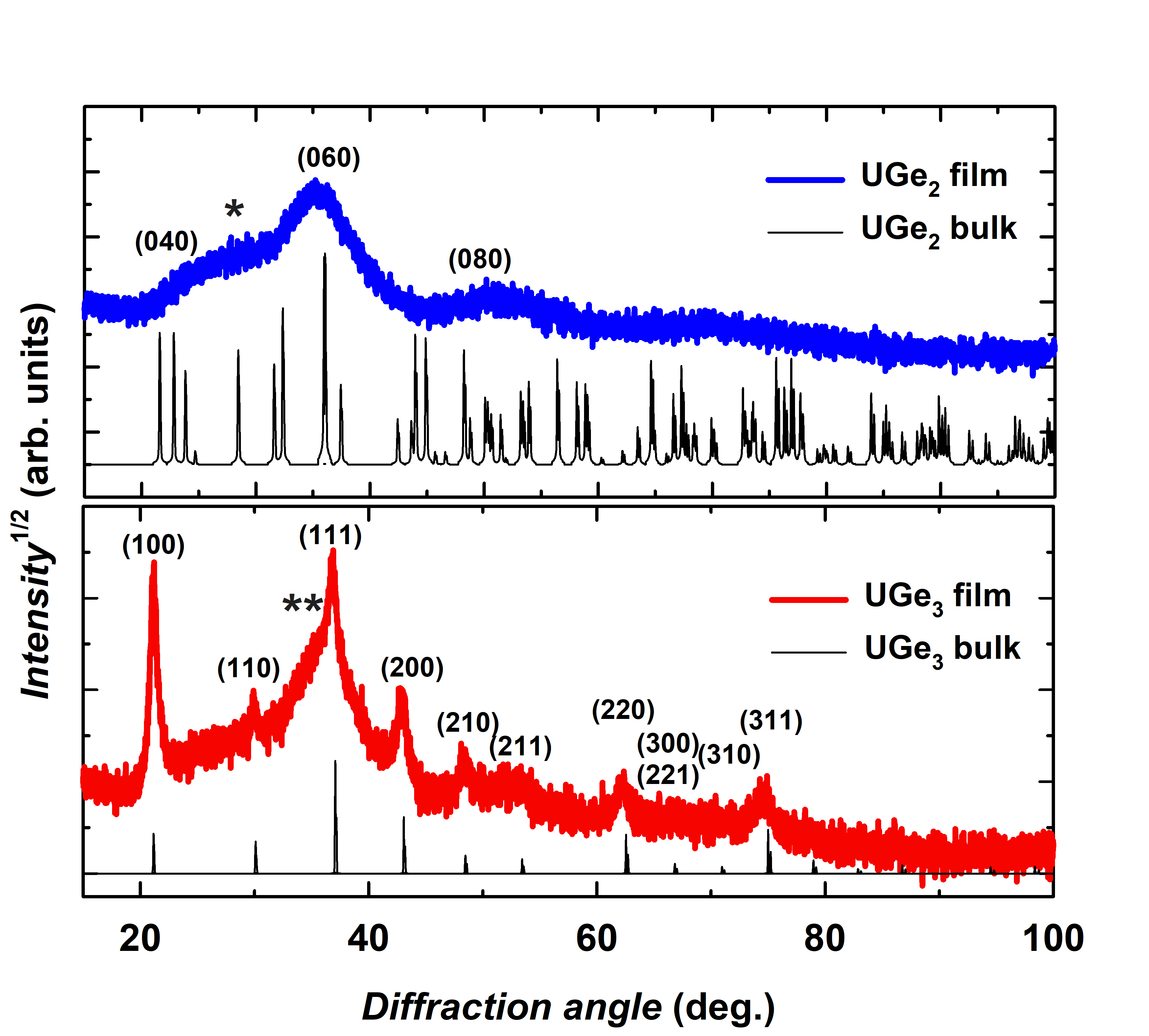}
    \caption{Grazing-incidence X-ray diffraction (GIXRD) patterns of UGe$_2$ (top) and UGe$_3$ (bottom) thin films measured with an incidence angle $\alpha$$_i$ = 0.6$^\circ$ using Cu K$\alpha$ radiation. Simulated bulk diffraction patterns are shown for comparison (black lines). }
    \label{fig:xps_survey}
\end{figure}

The UGe$_2$ film exhibits predominantly (0$k$0) reflections, indicating pronounced preferred orientation of the crystallites. The broad diffraction features and elevated background suggest nanocrystalline grain size and structural disorder, consistent with room-temperature preparation conditions. The diffuse feature marked by a single asterisk (*) may originate from the (111) reflection of UO$_2$. Although the samples contain less than 4\% oxygen when produced and tested \textit{in situ}, some oxidation may occur post deposition. The diffraction pattern of the Ge-rich film shows several reflections broadly consistent with the bulk UGe$_3$ structure. Compared to the UGe$_2$ film, the diffraction peaks are noticeably sharper, indicating improved crystallinity. Weak additional features marked by double asterisks (**) are not reproduced by the UGe$_3$ reference pattern and may indicate a minor secondary phase or structural disorder within the film.

\FloatBarrier
\clearpage
\section{The detailed U-4f core-level XPS study}
\FloatBarrier

\begin{figure}[!htbp]
    \centering
    \includegraphics[width=0.9\textwidth]{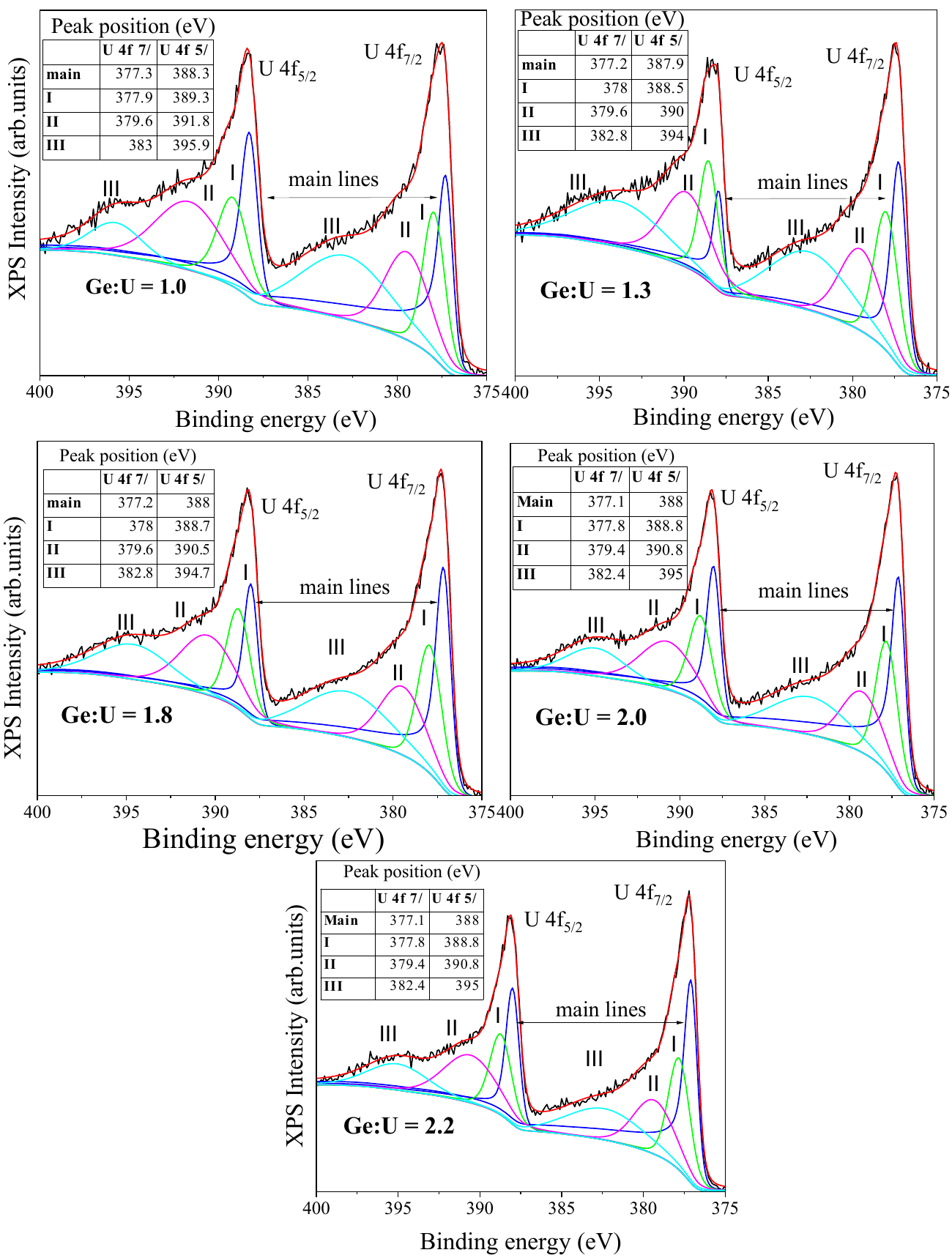}
    \caption{U 4f core-level XPS spectra of UGe$_{2\pm x}$ thin films for different compositions (Ge:U $=$ 1.0, 1.3, 1.8, 2.0, and 2.2) together with the corresponding peaks deconvolution. The experimental data (black) are shown with the total fit (red) and individual fitted components (colored). The fitting was performed using a constrained model where the main U~4$f_{7/2}$ and U~4$f_{5/2}$ lines (marked "main lines") were restricted to a fixed 3:4 intensity ratio. In addition to the main doublet, three additional doublets (I--III) were included to reproduce higher-binding-energy contributions. The extracted peak positions (binding energies) for U~4$f_{7/2}$ and U~4$f_{5/2}$ for the main lines and components I--III are listed in the inset tables for each Ge:U ratio.}

    \label{fig:xps_ge_deconvolution}
\end{figure}

\FloatBarrier
\clearpage
\section{The detailed Ge-3d core-level XPS study}
\FloatBarrier

\begin{figure}[!htbp]
    \centering
    \includegraphics[width=0.95\textwidth]{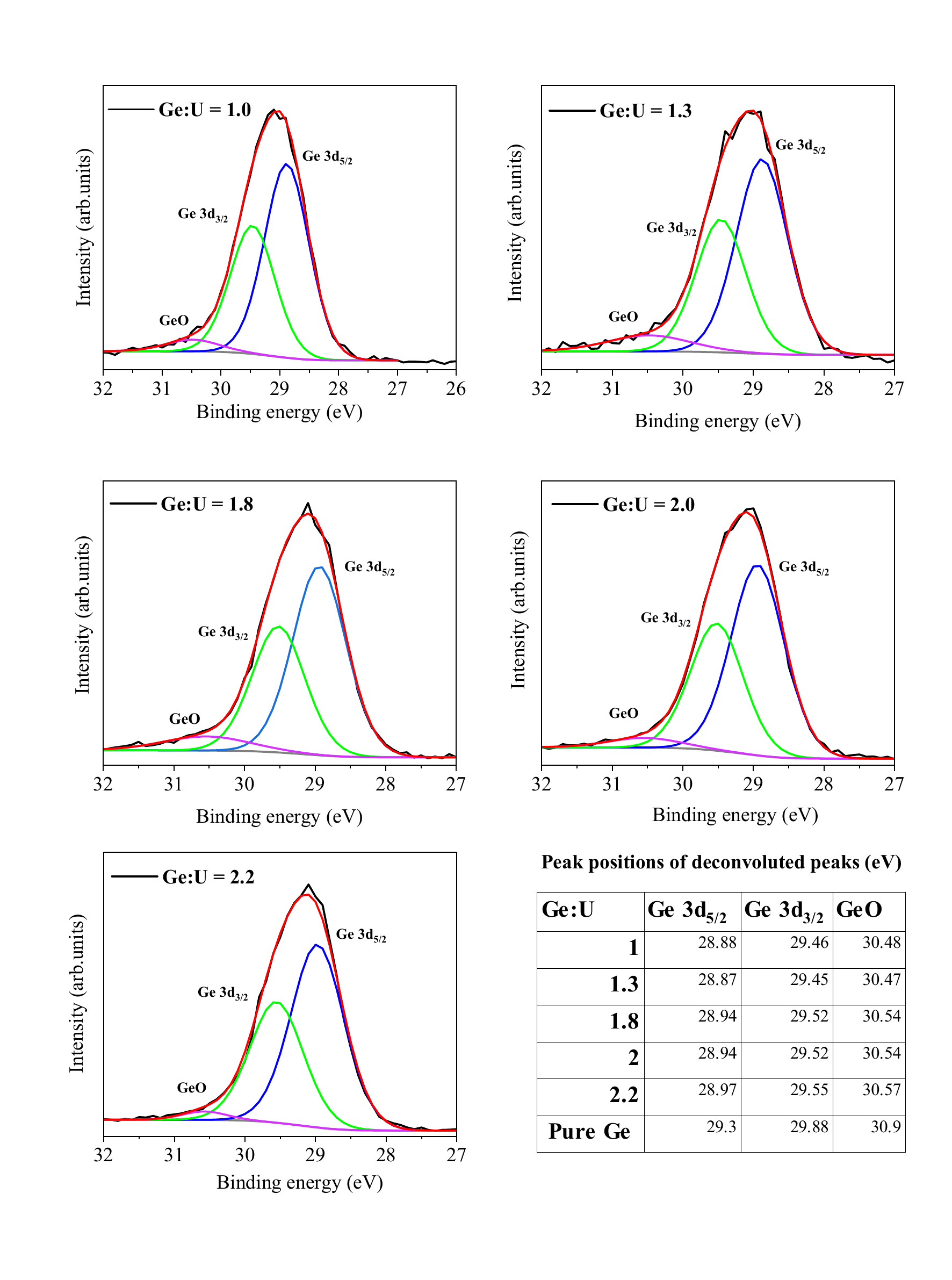}
    \caption{Ge~3d core-level XPS spectra of UGe$_2$ for different Ge:U ratios in UGe$_{2\pm x}$ thin films. 
    The experimental data are fitted using constrained Ge~3d spin--orbit doublets with a fixed 3d$_{5/2}$--3d$_{3/2}$ splitting of $\sim$0.58~eV, a statistical area ratio of 3:2, and identical line shapes. A weak high-binding-energy shoulder attributed to a minor GeO contribution (2–3\%) is present in all samples. The table summarizes the fitted peak positions.}
    \label{fig:xps_ge_deconvolution}
\end{figure}
\FloatBarrier

\FloatBarrier
\clearpage

\end{document}